# Terahertz spin currents and inverse spin Hall effect in thin-film heterostructures containing complex magnetic compounds

T. Seifert[1], U. Martens[2], S. Günther[3], M. A. W. Schoen[4], F. Radu[5], X. Z. Chen[6], I. Lucas[7] , R. Ramos[8], M. H. Aguirre[9], P. A. Algarabel[10], A. Anadón[9], H. Körner[4], J. Walowski[2], C. Back[4], M. R. Ibarra[7], L. Morellón[7], E. Saitoh[8], M. Wolf[1], C. Song[6], K. Uchida[11], M. Münzenberg[2], I. Radu[5], T. Kampfrath[1,*]

[1]Physical Chemistry, Fritz Haber Institute of the Max Planck Society, 14195 Berlin, Germany

[2]Institute of Physics, Ernst Moritz Arndt University, 17489 Greifswald, Germany

[3]Multifunctional Ferroic Materials Group, ETH Zürich, 8093 Zürich, Switzerland

[4]Institute for Experimental and Applied Physics, University of Regensburg, 93053 Regensburg, Germany

[5]Max-Born Institute for Nonlinear Optics and Short Pulse Spectroscopy, 12489 Berlin, Germany

[6]Key Laboratory of Advanced Materials, School of Materials Science and Engineering, Tsinghua University, 100084 Beijing, China

[7]Instituto de Nanociencia de Aragón, Universidad de Zaragoza, E-50018 Zaragoza, Spain

[8]WPI Advanced Institute for Materials Research, Tohoku University, 980-8577 Sendai, Japan

[9]Departamento de Física de la Materia Condensada, Universidad de Zaragoza, 50009 Zaragoza, Spain

[10]Instituto de Ciencia de Materiales de Aragón, Universidad de Zaragoza and Consejo Superior de Investigaciones Científicas, E-50009 Zaragoza, Spain

[11]National Institute for Materials Science, 305-0047 Tsukuba, Japan

* E-mail: kampfrath@fhi-berlin.mpg.de

Abstract: Terahertz emission spectroscopy of ultrathin multilayers of magnetic and heavy metals has recently attracted much interest. This method not only provides fundamental insights into photoinduced spin transport and spin-orbit interaction at highest frequencies but has also paved the way to applications such as efficient and ultrabroadband emitters of terahertz electromagnetic radiation. So far, predominantly standard ferromagnetic materials have been exploited. Here, by introducing a suitable figure of merit, we systematically compare the strength of terahertz emission from X/Pt bilayers with X being a complex ferro-, ferri- and antiferromagnetic metal, that is, Dysprosium Cobalt ($DyCo_5$), Gadolinium Iron ($Gd_{24}Fe_{76}$), Magnetite ($Fe_3O_4$) and Iron Rhodium (FeRh). We find that the performance in terms of spin-current generation not only depends on the spin polarization of the magnet's conduction electrons but also on the specific interface conditions, thereby suggesting terahertz emission spectroscopy to be a highly surface-sensitive technique. In general, our results are relevant for all applications that rely on the optical generation of ultrafast spin currents in spintronic metallic multilayers.

*Keywords*: terahertz spintronics, femtomagnetism, spin Hall effect, spin Seebeck effect, heterostructures

## 1. Introduction

Exploiting the electron's spin degree of freedom is envisioned to be of central importance for future information technology.[1] In spintronic devices, the building blocks are related to the efficient generation, transport and detection of spin currents. New fundamental effects are currently in the focus of spintronics research, for instance the spin-dependent Seebeck effect (SDSE)[2], the spin Seebeck effect (SSE)[3] and the inverse spin Hall effect (ISHE)[4]. On one hand, the SDSE/SSE describes, respectively, the generation of a spin current carried by conduction electrons/magnons along a temperature gradient in a magnetically ordered solid. On the other hand, spin-current detection can be accomplished by the ISHE which transforms a spin current into a transverse charge current in materials with strong spin-orbit coupling.

A promising approach for characterizing materials relevant for spintronic applications is terahertz (THz) spectroscopy[5] and, in particular, THz emission spectroscopy (TES). As recently shown[6,7], upon illumination of magnetic heterostructures with femtosecond near-infrared laser pulses, a combination of the SDSE/SSE and the subsequent ISHE gives rise to the emission of electromagnetic radiation with frequencies extending into the THz range (see Fig. 1a).

Besides such material-science-driven interest, spintronic heterostructures also show a large potential as efficient and broadband THz emitters. So far, standard ferromagnetic materials have predominantly been explored with TES of magnetic heterostructures, that is, Ni, Co, Fe and binary alloys thereof.[7,8,9,10,11] However, a relatively straightforward access to a much larger variety of magnetic materials is provided by TES.

Here, we explore a number of complex metallic compounds in terms of their THz emission response following fs laser excitation of FM/Pt bilayers. Guided by a simple model of THz emission, our comparative study is performed under conditions that allow fair comparison of the FM materials. We provide a theory that highlights the key parameters for THz emission from FM/NM bilayers (e.g. impact of thickness, conductivity, spin Hall angle). The studied magnetic compounds exhibit different types of magnetic ordering: ferrimagnetic Magnetite ($Fe_3O_4$), (anti)ferromagnetic iron rhodium (FeRh), and the ferrimagnetic alloys dysprosium cobalt ($DyCo_5$) and gadolinium iron ($Gd_{24}Fe_{76}$).

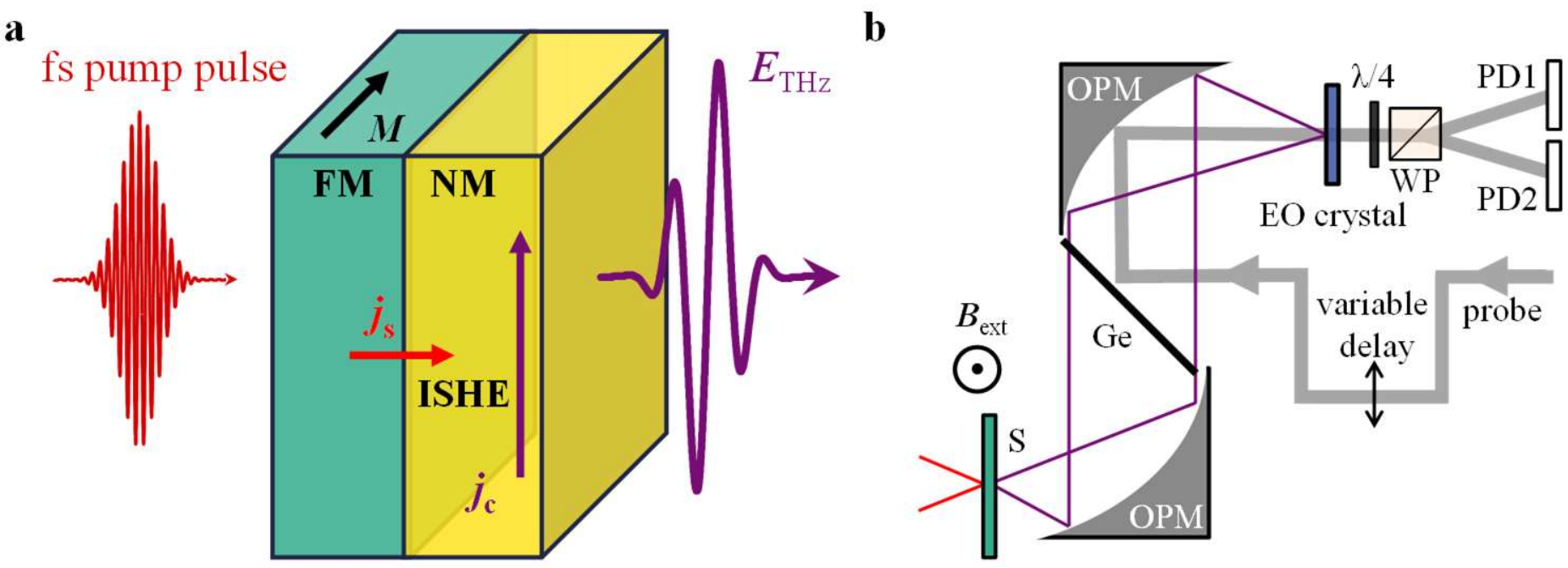

Figure 1. Experimental approach. **a**, Operational principle of the spintronic terahertz emitter. A femtosecond near-infrared pump pulse excites electrons throughout the nanometer-thick metallic heterostructure. Consequently, a spin current $\boldsymbol{j_s}$ is ejected from the in-plane magnetized ferromagnetic (FM) layer into the adjacent non-magnetic (NM) layer. Here, the inverse spin Hall effect (ISHE) converts the spin into a transverse charge current $\boldsymbol{j_c}$. This sub-picosecond in-plane charge current burst emits a terahertz pulse into the optical far-field. **b**, Schematic of the experimental setup. The near-infrared pump beam is focused onto the sample (S, in-plane sample magnetization is set by an external magnetic field $\boldsymbol{B_{ext}}$). The emitted terahertz beam is collimated and then focused again onto an electro-optic (EO) crystal with a pair of off-axis parabolic mirrors (OPM). A weaker femtosecond near-infrared probing pulse is overlapped in time and space with the THz pulse by means of a Germanium wafer (Ge) and a variable delay. A standard detection scheme consisting of a quarter-wave plate (λ/4), a Wollaston prism (WP) and two balanced photodiodes (PD1 and PD2) permits time-resolved detection of the THz electric field by delaying the two pulses with respect to one another.

## 2. Experimental details

### 2.1 *Principle*

A schematic of TES of magnetic heterostructures is depicted in Fig. 1a. A near-infrared femtosecond pump pulse drives the electronic systems of the ferri-/ferromagnetic (FM) and the non-magnetic (NM) layer out of equilibrium. Due to different transport properties for majority and minority spin channels in the FM material, an ultrafast spin current is launched into the NM layer. There, it is converted into a transverse charge current by means of the ISHE. Finally, this sub-picosecond in-plane charge current burst emits a THz electromagnetic pulse into the optical far-field. We note that the pump photon energy (~1.6 eV) leads to an electron distribution function that has strong nonequilibrium character directly after sample excitation.[12] Since relaxation toward a Fermi-Dirac distribution proceeds on a time scale of 100 fs,[13] both nonequilibrium and equilibrium electrons are expected to contribute to spin transport and THz emission.[14]

The emitted THz electric field directly behind the sample can straightforwardly be calculated in the limit of thin metal films (total metal thickness $d$ small compared to attenuation length and wavelength of the THz field within the metal). In the frequency domain, we obtain[7]

$$E_{\mathrm{THz}}(\omega) = C(\omega) \cdot j_{\mathrm{s}}^{0}(\omega)\gamma\lambda_{\mathrm{rel}} \qquad (1)$$

where $j_{\mathrm{s}}^{0}$ is the spin current density injected into the NM layer (measured directly behind the FM/NM interface and normalized to the pump excitation density), and $\gamma$ and $\lambda_{\mathrm{rel}}$ are the spin Hall angle and relaxation length of the THz spin current within the NM layer. The function

$$C(\omega) = \frac{A/d}{(n_1 + n_2)/Z_0 + \int_0^d dz\, \sigma(z)} \qquad (2)$$

quantifies how efficiently pump and THz radiation are, respectively, coupled into and out of the metal stack. $C$ summarizes all sample parameters unrelated to spintronic properties. In Eq. (2), $A$ is the pump pulse energy absorbed by the stack having thickness $d$, $n_1$ and $n_2$ are the refractive indices of air and the substrate, respectively, $Z_0 = 377\,\Omega$ is the vacuum impedance, and $\sigma$ is the conductivity at THz frequencies. As shown below, $C(\omega)$ depends on frequency only weakly and can approximately be considered as a constant.

It is important to note that $j_{\mathrm{s}}^{0}$ includes spin currents generated by the SDSE and the SSE as well as a spin-dependent FM/NM-interface transmission amplitude, which may depend on sample preparation details. In the derivation of Eq. (1), multiple reflections of the spin current inside the NM layer are neglected since all Pt layer thicknesses throughout this work are well above the relaxation length $\lambda_{\mathrm{rel}}$ of Pt (~1 nm).[7] The linear scaling of the THz electric field with the absorbed pump power in connection with Eq. (1) reflects the second-order nonlinearity of the spin-current generation process.

Maximizing the performance of the THz emitter thus requires the optimization of the material parameters. Previous work has shown that Pt with a thickness of about 3 nm is the best choice for the NM material.[7] Here, we focus on the variation of the FM material which has a direct impact on the magnitude of $j_{\mathrm{s}}^{0}$ and $C$ (through $A$ and $\sigma$). To quantify the efficiency of a magnetic material X in injecting a spin current into the adjacent Pt layer in a X/Pt bilayer, we introduce a figure of merit (FOM) that compares the normalized THz emission strength from a bilayer X/Pt to that of a CoFeB/Pt reference sample. The FOM is calculated according to

$$\mathrm{FOM_X} = \frac{\left\|j_{\mathrm{s,X}}^{0}\right\|}{\left\|j_{\mathrm{s,ref}}^{0}\right\|} = \frac{\|S_{\mathrm{X}}\|/C_{\mathrm{X}}}{\|S_{\mathrm{ref}}\|/C_{\mathrm{ref}}} \qquad (3)$$

where $\|S\|$ denotes the maximum magnitude of the measured THz signal waveform $S(t)$. We choose CoFeB as the reference ferromagnet because it features the highest THz emission performance among Ni, Co, Fe and their binary alloys.[7] To summarize, our macroscopic model of the THz emission amplitude [Eqs. (1), (2), (3)] allows us to perform a systematic comparison of different magnetic materials X and to disentangle the crucial material parameters to obtain maximum THz emission.

### 2.2 *Setup*

A schematic of the experimental setup is shown in Fig. 1b. The sample under study is pumped with laser pulses from a Ti:sapphire laser oscillator (duration 10 fs, center wavelength 800 nm, pulse energy 2.5 nJ, repetition rate 80 MHz). Note that some samples (see Table 1) have an unpolished substrate backside and are, therefore, studied in reflection geometry under an pump-beam angle of incidence of 45°. The transient field of the emitted THz pulse is detected via the linear electrooptic effect by a co-propagating probe pulse from the same laser (pulse energy 0.6 nJ) in a standard 1 mm thick (110)-oriented ZnTe electrooptic crystal.[15]

The sample magnetization is saturated in the film plane by means of two permanent magnets. For measurements in transmission and reflection geometry, we apply a magnetic field of ±70 mT and ±100 mT, respectively. By switching between opposite magnetization directions, the contribution odd in the sample magnetization can be extracted. For the samples studied here, the contribution even in the magnetization is typically one order of magnitude smaller and is neglected throughout this article. The temperature of the sample is varied by thermal contact with a massive metal block that is either cooled by a liquid nitrogen reservoir or heated by a resistive coil attached to it. To avoid water condensation, we apply a steady flow of gaseous nitrogen directed onto the sample surface. During temperature-dependent measurements, a magnetic field of 40 mT is applied.

The absorptance of the near-infrared pump pulse is determined by measuring the power reflected by and transmitted through the sample. To determine the THz conductivity of the thin film, we perform THz transmission measurements, referenced to a part of the substrate free of any sample material.[16] All THz measurements are conducted in a dry nitrogen atmosphere.

### 2.3 *Samples*

The studied samples consist of two or three metal layers. The bilayer structure is X/Pt with X being the magnetic compound, while Pt is chosen as the nonmagnetic layer because of its large spin Hall angle. The trilayer structure is Pt/X/Pt which allows for a consistency check of our theoretical model (see Fig. 1a). An overview of all samples used for THz emission measurements is given in Table 1. Details on sample fabrication can be found in Appendix A.

| # | Sample structure | Prepared by | AOI (°) | Absorptance | $C$ ($10^9$ Ω/m) | FOM |
|---|---|---|---|---|---|---|
| 1 | Sapphire//$DyCo_5$(3)/Pt(3) | Berlin | 0 | 0.56 | 3.5 | 0.85 |
| 2 | Sapphire//$Gd_{24}Fe_{76}$(3)/Pt(3) | Berlin | 0 | 0.52 | 3.0 | 0.79 |
| 3 | Sapphire//Pt(3)/$DyCo_5$(3)/Pt(3) | Berlin | 0 | 0.56 | 1.5 | - |
| 4 | Sapphire//Pt(3)/$Gd_{24}Fe_{76}$(3)/Pt(3) | Berlin | 0 | 0.53 | 1.3 | - |
| 5 | Glass//CoFeB(3)/Pt(3) | Greifswald | 0 | 0.57 | 3.5 | 1.00 |
| 6 | MgO//$Fe_3O_4$(24)/Pt(8) | Zaragoza | 45 | 0.44 | 2.7 | 0.09 |
| 7 | MgO//CoFeB(25)/Pt(5) | Greifswald | 45 | 0.52 | 2.5 | 1.00 |
| 8 | MgO//$Fe_{51}Rh_{49}$(10)/Pt(5) | Beijing | 45 | 0.60 | 1.0 | 0.28 |
| 9 | Glass//CoFeB(10)/Pt(2) | Greifswald | 45 | 0.47 | 1.2 | 1.00 |
| 10 | Glass//CoFeB(20)/Pt(2) | Greifswald | 45 | 0.46 | 0.5 | 1.00 |

Table 1. Overview of samples used for terahertz emission measurements. Sample structure including location of preparation. Also given are the pump beam angle of incidence (AOI), absorptance of the near-infrared pump light, the coupling function $C$ (calculated at a frequency of 1 THz, see Eq. (2)) and the figure of merit [FOM, see Eq. (3)]. In the column "Sample structure", the numbers in brackets indicate the film thickness in nm.

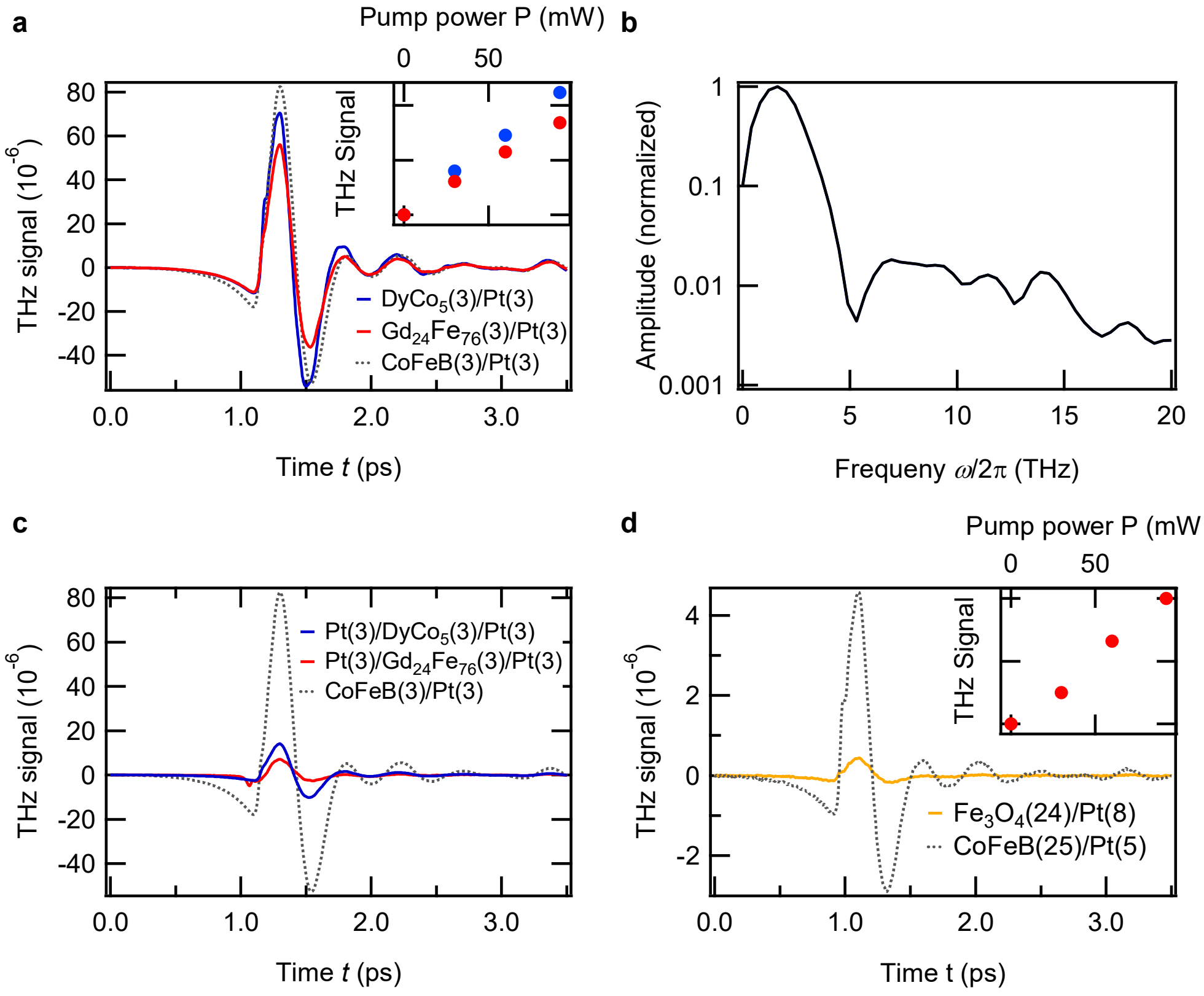

Figure 2. **Terahertz emission from complex magnetic compounds.** Raw data comparing the terahertz emission (odd in sample magnetization) from magnetic heterostructures containing **a**, **b**, $DyCo_5$ and $Gd_{24}Fe_{76}$, **c**, symmetric trilayer structures and **d**, $Fe_3O_4$. The dashed line of each panel shows emission from a CoFeB/Pt reference bilayer with similar structure. **Insets,** Pump power dependence of the respective THz signal (RMS). The amplitude of the Fourier-transformed waveform emitted by the bilayer containing $DyCo_5$ is shown in panel **b**.

To evaluate each FM material's capability to emit a spin current into an adjacent NM layer according to Eq. (3), we fabricate for each of the three sample groups (see Table 1) a suitable CoFeB/Pt reference sample of similar structure. These reference samples have coupling functions $C$ [see Eq. (2) and Table 1] comparable to those of their counterparts containing complex magnetic compounds.

## 3. Results and discussion

### 3.1 *$DyCo_5$ and $Gd_{24}Fe_{76}$*

Metallic ferrimagnetic alloys consisting of rare-earth (RM) and transition metal (TM) elements, such as $DyCo_5$ and $Gd_{24}Fe_{76}$, have been among the first magnetic media used for high-density magnetooptical recording.[17] Because of their strong perpendicular magnetic anisotropy, tunable magnetic properties, large magnetooptical effects and, consequently, enlarged signal-to-noise ratios in magnetooptical detection (due to their amorphous state), they found applications as the first magnetic re-writable memories. More recently, the all-optical magnetization switching phenomenon (i.e. purely laser-driven spin switching in the absence of an external magnetic field) has been discovered on these ferrimagnets,[18,19,20] which has brought this class of RM-TM alloys into the focus of ultrafast magnetic studies over the last years.[21]

Figure 2 shows typical THz waveforms emitted from magnetic heterostructures containing $DyCo_5$ and $Gd_{24}Fe_{76}$ (panels a, b, and c). For comparison, we also show the THz waveform from the CoFeB/Pt reference sample that has a similar thickness (dashed lines in Figs. 2a and c, for sample details see Table 1). For all these samples, we observe similar temporal dynamics of the THz signal waveform and a linear pump power dependence (see inset of Fig. 2a), as expected for a second-order nonlinear process. Fourier-transformation of the emitted waveform from the bilayer containing $DyCo_5$ yields the complex-valued spectrum whose amplitude is shown in Fig. 2b. We note that the bandwidth of about 3 THz is limited by the 1 mm thick ZnTe detection crystal used for EO sampling.

The DC conductivities $\sigma$ as determined by THz transmission spectroscopy[22] are $3.1\cdot 10^5$ S/m for $DyCo_5$, $9.0\cdot 10^5$ S/m for $Gd_{24}Fe_{76}$, $13.5\cdot 10^5$ S/m for CoFeB and $50.5\cdot 10^5$ S/m for Pt. These values are approximately constant up to 10 THz. Thus, we find an almost frequency-independent and similar coupling function $C$ for samples containing $DyCo_5$, $Gd_{24}Fe_{76}$ and the respective reference (compare Table 1). This fact enables direct comparison of the raw data in Figs. 2a and c.

We observe that heterostructures containing $DyCo_5$ show similar THz emission and, thus, magnitude of $j_s^0$ as the reference ferromagnetic CoFeB/Pt sample. This result is remarkable given the reduced net magnetization due to the ferrimagnetic order of $DyCo_5$. In contrast, $Gd_{24}Fe_{76}$-capped heterostructures show reduced performance as spin ejector compared to CoFeB/Pt layers. In terms of the FOMs, we find 0.85 for $DyCo_5$ and 0.79 for $Gd_{24}Fe_{76}$ (compare with Eq. (3) and Table 1).

Such lowered performance could arise from a reduced spin polarization of conduction electrons which are believed to play a key role in the spin-current generation. Although no published data on the spin polarization for $DyCo_5$ and $Gd_{24}Fe_{76}$ are available, we estimate it to be about 40% and 36%, respectively, based on Eq. (S1) in the supplementary information of Ref. 23. Note that this estimation neglects any contribution from 5d electrons. These spin-polarization values are lower than the reported values[24] for CoFeB (~65%) and could, thus, explain the observed differences in terms of spin-current emission.

This reasoning is further supported by the relatively low Curie temperature of 550 K for $Gd_{24}Fe_{76}$ as compared to 925 K for $DyCo_5$ (Ref. 25). At room temperature, the lower critical temperature might lead to a reduction of the relative spin polarization of the conduction electrons and, in turn, to a reduced spin current. This effect can be enhanced by accumulative heating of the sample region at the laser focus by the train of pump pulses (repetition rate is 80 MHz).

### 3.2 *Symmetric trilayers*

As a check of the sample quality, we also conducted measurements on samples having the symmetric structure Pt/X/Pt, in which the FM layer X=$DyCo_5$ and $Gd_{24}Fe_{76}$ is embedded between two Pt layers of nominally identical thickness (see Table 1). As seen in Fig. 2c, the THz emission measurements on the symmetric trilayers yield a THz signal amplitude about one order of magnitude lower than for the respective bilayers. This behavior can be understood based on our picture of the microscopic mechanism underlying THz emission from magnetic heterostructures (see Fig. 1a and Ref. 7). Since the pump field is homogeneous throughout the thickness of the thin-film sample, the backward- and forward-directed spin currents injected into the back and front Pt layer generate transverse charge currents that cancel each other. Consequently, the resulting THz emission is quenched, consistent with our experimental observation and indicating a high sample quality. The small residual emission may originate from a slight sample asymmetry, for example due to slightly different Pt film thicknesses or higher strain closer to the substrate.

### 3.3 *$Fe_3O_4$*

Magnetite ($Fe_3O_4$) is one of the strongest naturally occurring ferrimagnets and shows a Verwey transition at a temperature of typically 120 K.[26] The spin polarization at the Fermi energy is predicted to be close to unity, which makes this material promising for spintronic applications.[27]

The $Fe_3O_4$ sample shows a THz emission that is about 10 times smaller than from the CoFeB/Pt reference (Fig. 2d). The THz signal amplitude again depends linearly on the pump power as seen from the inset of Fig. 2d and the THz waveform shows similar temporal dynamics as the reference. Magnetite's DC conductivity of about $0.1\cdot10^5$ S/m (Ref. 28) is about two orders of magnitude lower than that of CoFeB. However, due to a different NM/FM-layer thickness ratio of the Magnetite sample, the coupling function $C$ has a magnitude similar to that of the reference sample (see Table 1). Note that any inhomogeneity in the excitation density across the metal stack is expected to be balanced out within a few tens of femtoseconds due to electron transport.[29] We extract a FOM for $Fe_3O_4$ of 0.09 (see Eq. (3) and Table 1).

The low efficiency of spin current emission of $Fe_3O_4$ into Pt cannot straightforwardly be understood in terms of its spin polarization since $Fe_3O_4$ is believed to be a half-metal with reported experimental spin polarization values[30] of about 72%, which is larger than that of CoFeB (~65%).[24] Nonetheless, it is well known that Magnetite's room-temperature conductivity is governed by electron hopping[31,32] and much lower than for CoFeB. On one hand, this conduction mechanism could diminish the SDSE contribution to the spin current and so affect the efficiency of spin current emission into an adjacent Pt layer, in agreement with previous spin pumping experiments[33]. On the other hand, the SSE contribution to the spin current is believed to be much weaker than the SDSE in general, as indicated by measurements in bilayers containing a FM insulator (not shown).

### 3.4 *FeRh*

Iron Rhodium is a remarkable material as it exhibits a transition from an antiferromagnetic (AFM) to a FM phase at a temperature that is strongly depending on the exact composition and sample preparation.[34,35] This feature makes FeRh a promising candidate for heat-assisted magnetic recording, which benefits from the inherent magnetic stability of the AFM ordering.[36]

Figure 3 shows temperature-dependent THz emission data from an FeRh sample (15 nm total metal film thickness) in comparison to two CoFeB/Pt reference samples (12 and 22 nm total metal film thicknesses). For all these samples, we again observe similar temporal dynamics of the emitted THz waveform. We find that the FeRh sample has a lower THz emission performance than both CoFeB/Pt reference samples at room temperature (Fig. 3a).

Remarkably, the emitted THz signal scales quadratically with the pump power at 300 K (see inset of Fig. 3a), whereas it scales linearly at an elevated temperature of 350 K. From measurements employing a superconducting quantum interference device (SQUID), we find that the studied FeRh sample undergoes a transition from an antiferromagnetically to a ferromagnetically ordered state at ~320 K (see Fig. 3b). Therefore, the distinct pump power dependencies might be due to the mixed AFM/FM state of the FeRh sample just above room temperature. In this regime, the magnetization and, thus, spin polarization scales roughly linearly with temperature. Therefore, the pump pulse plays a two-fold role: it not only heats the electrons transiently (linear absorption) but it also increases the sample temperature statically via accumulative heating by many pump pulses. This double action explains the observed quadratic pump power scaling at room temperature. Note that any significant contribution of a single laser pulse to driving the phase transition is unlikely because the pump pulse fluence (about 0.1 mJ/cm$^2$) is ten times smaller than the critical fluence found in previous pump-probe works on comparable samples.[37]

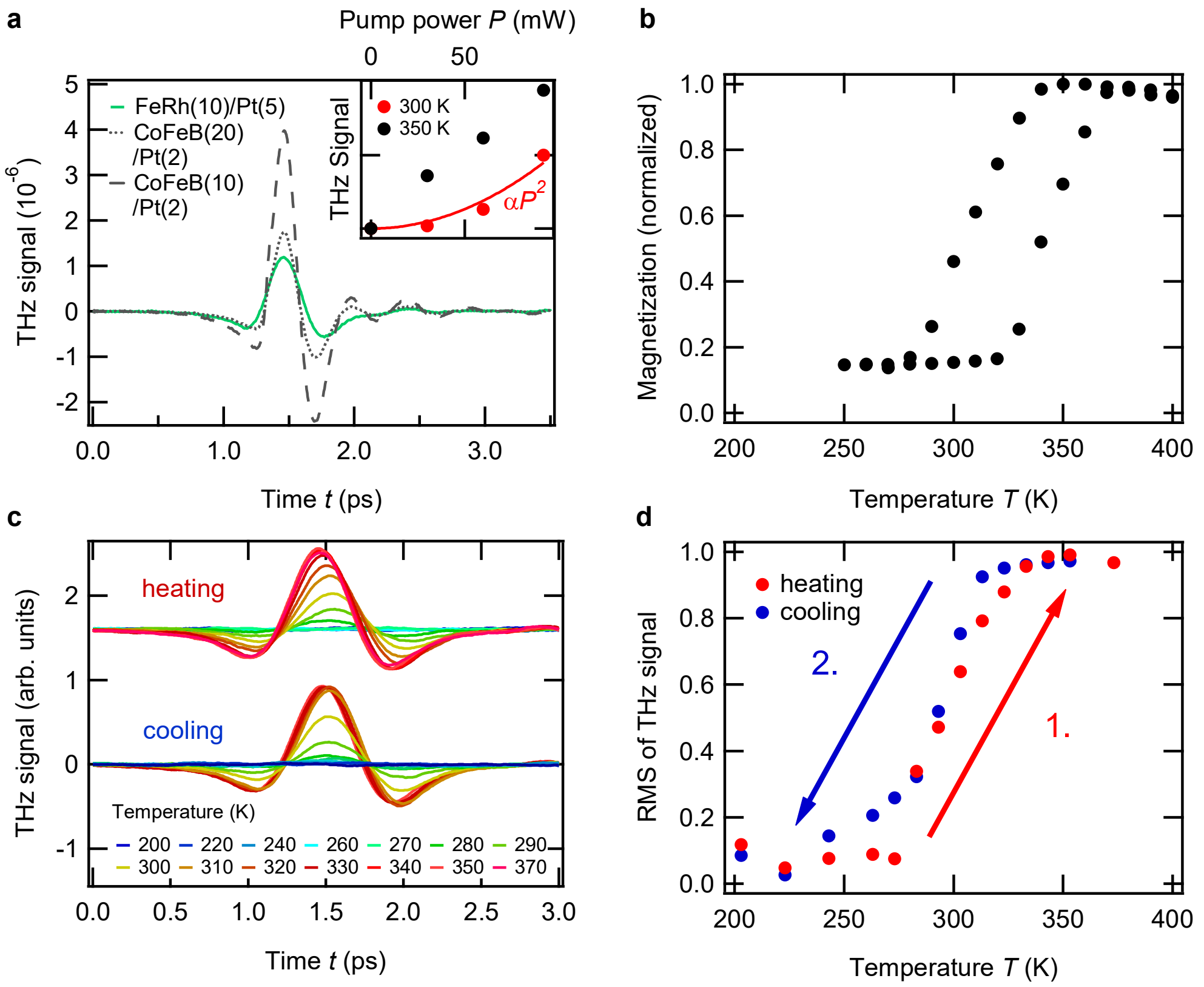

Figure 3. **Temperature-dependent terahertz emission from FeRh/Pt. a,** Raw data comparing the terahertz emission (odd in sample magnetization) at room temperature from a magnetic heterostructure containing FeRh with two CoFeB/Pt bilayers having similar thicknesses (dashed lines). **Inset,** Pump power dependence of the THz signal (RMS) at room temperature (red dots) together with a quadratic fit (solid red line) and at 350 K (black dots). **b,** Temperature dependence of the sample bulk magnetization measured by SQUID. **c,** Dependence of the terahertz emission on sample temperature below and above the antiferromagnet-to-ferromagnet transition temperature. **d,** Temperature dependence of the root mean square (RMS) of the terahertz signal.

From the DC conductivity of FeRh ($3.3 \cdot 10^5$ S/m, Ref. 38), we deduce a coupling function $C$ comparable to that of the 12 nm thick CoFeB/Pt reference sample, whereas $C$ is twice as large as for the 22 nm thick reference sample (see Eq. (2) and Table 1). Since even the 22 nm thick CoFeB/Pt sample shows higher THz emission efficiency than the FeRh film (despite its lower coupling function $C$), we conclude that FeRh is a less efficient spin-current emitter than CoFeB at room temperature and for the applied pump fluence. This notion is bolstered by the calculated FOM of 0.28 for FeRh relative to the reference samples [see Eq. (3) and Table 1]. However, the pump-power dependence of the THz signal amplitude at elevated temperatures (see inset of Fig. 3a) suggests a two-fold enhancement of the THz signal amplitude, potentially approaching the performance of CoFeB, at significantly higher pump powers than utilized in this study. It is noteworthy that the two reference samples exhibit identical FOMs despite their different thickness, thereby demonstrating the robustness of our evaluation method.

We also perform temperature-dependent THz emission measurements on FeRh (see Figs. 3c and d). As shown in Fig. 3c, we observe a complete quenching of the THz emission signal upon cooling the sample below 250 K. When the sample is subsequently heated back to room temperature and above, the THz signal completely recovers. The temperature dependence of the root mean square of the THz signal is displayed by Fig. 3d, demonstrating the reversibility of the AFM/FM phase transition. Note, that the temperature range in our experiment should suffice to fully set the FeRh into the AFM and FM state, respectively (see SQUID measurements in Fig. 3b).

Interestingly and in contrast to the SQUID measurements, we do not observe a clear hysteretic behavior in our THz data. A similar phenomenon has been observed in previous experiments.[39,40] Instead, the THz signal amplitude rather seems to follow the cooling branch of the SQUID hysteresis with indications of a small hysteretic behavior at the kink regions (around 270 and 310 K).

We note that the nominal temperatures in Figs. 3a, c and d refer to the substrate temperature and not the actual sample temperature at the focus of the laser beam which is increased due to accumulative pump heating. Based on the pump power dependence and the SQUID data, we estimate this temperature discrepancy to be below 20 K. However, such accumulative heating of the sample would just lead to a rigid shift of the equilibrium hysteresis loop (Fig. 3b) along the temperature axis, in contrast to the temperature dependence of the THz signal amplitude observed here (Fig. 3d).

On the other hand, a single pump pulse transiently increases the electronic temperature by several 100 K. Thus, a second potential explanation for the distinct temperature hysteresis is that the pump pulses transiently lower the magnetic domain nucleation energy barrier, thereby shrinking the THz temperature hysteresis close to the critical temperature. This notion is bolstered by the experimentally observed small hysteretic behavior in the kink regions further away from the critical temperature, where the nucleation energy barrier could not yet be sufficiently lowered by the pump pulses.[41]

A third possible scenario may be related to the fact that the magnetic structure of the FeRh sheet close to the Pt interface is modified as observed in earlier studies.[39,42,43] Along these lines, it has been shown[7] that the laser-induced ultrafast spin currents decay within a length $\lambda_{\mathrm{rel}}$ of about 1 nm in Pt (see Eq. (1)). We anticipate similar length scales inside the FM layer. This fact suggests that TES of magnetic heterostructures is in general more sensitive to the interfacial region between FM and NM layers than to their bulk. This interpretation is plausible because at least at 350 K, the THz signal depends quadratically on the pump field (i.e. the laser power) and must, therefore, to a large extent arise from photoinduced THz currents flowing in regions with broken inversion symmetry, thus, close to the interface of the thin film studied here. Consequently, the above-mentioned differences between temperature-dependent SQUID and THz emission measurements might be also understood in terms of an altered surface magnetism in FeRh. This notion is bolstered by the remarkable agreement with the results obtained in Ref. 39.

## 4. Conclusion

In conclusion, we demonstrate the feasibility of terahertz emission spectroscopy in conjecture with complex magnetic metallic compounds. We introduce a figure of merit that permits direct comparison of the spin injection efficiency of different magnetic materials into an adjacent layer. This efficiency is not only relevant for the development of better spintronic THz emitters, but for all research involving ultrafast spin current injection, including spin control by the spin transfer torque.[12]

We find that X=CoFeB is still the most efficient spin current emitter in X/Pt-type bilayers. The observed differences in THz emission performance between the various magnetic materials may be understood in terms of the spin polarization at the Fermi energy for samples containing $DyCo_5$ and $Gd_{24}Fe_{76}$. However, our data on $Fe_3O_4$ indicate a crucial role of the particular conduction mechanism and the spin-dependent FM/NM interface transmission. Our results on FeRh further suggest that terahertz emission spectroscopy provides additional insights into the magnetic structure of a broad range of materials compared to well-established techniques such as SQUID. Further experiments involving half-metallic or spin-gapless semiconductors may help clarify the role of (non-)thermal electrons during the THz emission process.

Finally, the few-nanometer length scale over which the THz currents flow across the interface might ultimately lead to a sensitive probe of surfaces and buried interfaces. To further explore this opportunity, future studies with a profound control of interface parameters are required.

**Acknowledgements**

This work was supported by ERATO "Spin Quantum Rectification" and PRESTO "Phase Interfaces for Highly Efficient Energy Utilization" from JST, Japan, Grant-in-Aid for Scientific Research (A) (JP15H02012), Grant-in-Aid for Scientific Research on Innovative Area "Nano Spin Conversion Science" (JP26103005) from JSPS KAKENHI, Japan. The authors acknowledge support by the Spanish Ministry of Economy and Competitiveness (through Project Nos. MAT2014-51982-C-R, including FEDER funding) and the Aragon Regional government (Project No. E26). I.R. acknowledges funding from German Ministry for Education and Research (BMBF) through project 05K16BCA Femto-THz-X. T. S. and T. K. acknowledge funding through the ERC H2020 CoG project TERAMAG/grant no. 681917, through EU FET Open RIA Grant no. 766566 (ASPIN) and through the DFG priority program SPP 1538/SpinCaT.

## Appendix

### A. *Sample preparation*

A.1 *Beijing: FeRh*

The FeRh film was grown on a (001)-oriented single crystal MgO substrate using DC magnetron sputtering. The base pressure of the chamber was $2\cdot10^{-5}$ Pa. The substrate was kept at 573 K for 30 min. Then, FeRh (thickness of 10 nm) was deposited at an Ar pressure of 0.7 Pa, corresponding to a stoichiometric $Fe_{51}Rh_{49}$ film. The sputtering power was 30 W for 3-inch-diameter $Fe_{50}Rh_{50}$ targets. Afterwards, the film was heated to 1023 K and annealed for 100 min. When the film had cooled down to room temperature, it was capped with 5 nm Pt in situ.[44]

A.2 *Helmholtz-Zentrum Berlin: $DyCo_5$ and $Gd_{24}Fe_{76}$*

Thin films (thickness of 3 nm) of ferrimagnetic amorphous $Gd_{24}Fe_{76}$ and polycrystalline $DyCo_5$ alloys were grown by magnetron sputtering on (11-20)-oriented single-crystal $Al_2O_3$ substrates at room temperature in an ultra-clean Ar atmosphere of $1.5\cdot10^{-3}$ mbar pressure. Pt films (thickness of 3 nm) were used as a capping layer. Alternatively, samples with both Pt capping and buffer layers (thickness of 3 nm) were grown. The stoichiometry of the ferrimagnetic alloy was controlled by varying the deposition rate of the separate elemental targets during the co-sputtering process.[45]

A.3 *Zaragoza: $Fe_3O_4$*

The $Fe_3O_4$ film (thickness of 24 nm) was grown on a (001)-oriented MgO substrate by pulsed laser deposition using a KrF excimer laser (248 nm wavelength, 10 Hz repetition rate, $3\cdot10^9$ W/cm$^2$ irradiance) in an ultrahigh-vacuum (UHV) chamber. The Pt film (thickness of 8 nm) was deposited in the same UHV chamber by DC magnetron sputtering without breaking the vacuum. Further details on the growth can be found in Ref. 46. The film thickness was measured by x-ray reflectivity and its structural quality was confirmed by x-ray diffraction and transmission electron microscopy. The film cross sections were prepared by focused ion beam and measured by high-angle annular dark-field scanning transmission electron microscopy. The measurements were carried out in a probe-aberration corrected FEI Titan 60-300 operated at 300 kV.

A.4 *Greifswald: CoFeB*

The samples were deposited on glass with a surface roughness <1 nm and (100)-oriented MgO substrates, both with the dimensions 10 × 10 × 0.5 mm$^3$. The amorphous CoFeB layers were fabricated by magnetron sputtering from a nominal target composition of $Co_{20}Fe_{60}B_{20}$. A detailed analysis yielded a Cobalt-Iron ratio of 32:68. The Pt films on top were deposited using electron-beam evaporation under UHV conditions with a base pressure of $5\cdot10^{-10}$ mbar after the sputtering procedure without breaking the vacuum. All substrates were kept at room temperature during the deposition.